\documentstyle[twocolumn,pra,aps]{revtex}
\begin{document}
\tightenlines


\title{\Large \bf
Asymptotic channels and  gauge
transformations of the time-dependent
Dirac equation for extremely relativistic heavy-ion collisions
}
\author{J. C. Wells$^{1,2}$, B. Segev$^2$, and J. Eichler$^3$}
\address{
    $^1$Center for Computational Sciences, Oak Ridge National Laboratory 
        Oak Ridge, Tenneessee, 37831-6203, USA }
\address{
    $^2$Institute for Theoretical Atomic and Molecular Physics, 
        Harvard-Smithsonian Center for Astrophysics, \\
        60 Garden Street, Cambridge, Massachusetts 02138, USA}
\address{
    $^3$Bereich Theoretische Physik, Hahn-Meitner-Institut Berlin,
        14109 Berlin, Germany
        and \\
        Fachbereich Physik, Freie Universit\"at Berlin, 14195 Berlin,
Germany }
\date{\today}
\maketitle
\begin{abstract}
We discuss the two-center, time-dependent Dirac equation
describing the dynamics of an electron during a peripheral,
relativistic heavy-ion collision at extreme energies.
We derive a factored form, which is exact in the high-energy limit,
for the asymptotic channel solutions of the Dirac equation,
and elucidate their close connection with   gauge
transformations which transform the dynamics into a representation
in which the
interaction between the electron and a distant ion is of short range.
We describe the implications of this relationship for solving the
time-dependent Dirac equation for extremely relativistic collisions.
\end{abstract}

{\em PACS number}: 34.50.-s, 25.75.-q, 11.80.-m, 12.20.-m


\section{Introduction}

Particle production via electromagnetic processes in peripheral
collisions of relativistic heavy ions has received significant
study recently, both experimentally
\cite{VD92,BG93,Ba94,VD94,BG94,DE95,CB97,BG97,VD97,KV98,BC98}
and theoretically (for reviews, see \cite{BB88,Ei90,EM95}),
due to anticipated experimental opportunities at
colliding-beam accelerators, and the importance of this
phenomena for the operation and performance of such facilities.
Also of interest is the opportunity to study strong-field QED
effects in particle production over a wide range of
charge and collision energy
\cite{greiner,workshop,Ba90b,RW91,BG92,IE93,Io94,HT95a,GW95,HT95b,AK97} .

The high-energy limit of peripheral
relativistic heavy-ion collisions has been recently examined,
and closed-form expressions for the amplitudes describing
electron-positron pair production have been obtained
\cite{Ba97,SW98,BM98,SW98b}.
These new results offer significant insight into 
the understanding of relativistic heavy-ion collision
dynamics \cite{Ba97,SW98,BM98,SW98b,ER98a,ER98b,HT98}.
In these works, the consequences of allowing the collision
velocity to approach the speed of light, i.e.\
$\beta^{} \equiv v/c \rightarrow 1$, and thus the collision
energy to approach infinity,
$\gamma \equiv (1-\beta^2)^{-1/2} \rightarrow \infty$,
have been investigated.
This limit has been motivated by the progress
toward new colliding-beam heavy-ion accelerator facilities
currently in various stages of construction and planning.
The Relativistic Heavy-ion Collider (RHIC) at Brookhaven
National Laboratory will begin operation in 1999, offering
collision velocities in the collider frame of
$\beta_C \approx 0.9999$.
The Large Hadron Collider (LHC), currently being planned at CERN, will
offer collision velocities which more closely approach the
speed of light, $\beta_C \approx 0.9999999 $.
Indeed, in experiments recently performed at CERN's SPS
\cite{VD92,VD94,VD97,KV98}, in which heavy-ions collide in a
fixed-target mode, the equivalent collider-frame collision
velocity exceeds  $0.99c $, suggesting that the high-energy
limit is already a meaningful and relevant approximation
for use in interpreting the experimental results \cite{SW98b}.

Of central importance to recent investigations of
the high-energy limit is the use of a simplified form, accurate
to leading order in the small parameter $\gamma^{-2}$, for the
Lorentz-boosted Coulomb potential \cite{RR84,Ja92,Ba95,SW98} acting between
the active electron and a bare nucleus. In this form, the dependence of the
interaction on the transverse electron coordinates separates from the
dependence on the longitudinal coordinate $z$ and the time $t$.
Moreover, the dependence on the latter arises in combinations
identified as the lightfront variables,
e.g.\ $ \tau_{\pm} \equiv (z \pm t)/2$,
in the form of a zero-range or {\it sharp} potential.
The separability of this interaction in the time-dependent,
two-center Dirac equation allows for its closed-form solution
\cite{Ba97,SW98,BM98,SW98b}.
However, this useful form becomes apparant 
at the high energy limit only after
applying phase transformations so as to remove the long-range
$z$ dependence of the interaction\cite{Ba95,BRW,BRW93}.

In this present work, we study
these phase transformations and show how they
constitute well-defined gauge transformations while from a parallel
perspective they
formally define an
interaction-representation in which
the asymptotic (i.e.\ $|t| \rightarrow \infty$)
interaction of an electron with a distant ion is
absorbed into a redefinition of the electronic states.
In this representation, which we call the {\it short-range
representation}, the asymptotic channel states are
free from effects of the distant ion,
and in the high energy limit of infinite
$\gamma$ the interaction has zero range.
In the high-energy limit, the separation is exact.
For finite $\gamma$, the short-range interaction is an approximation
correct to order $\gamma^{-2}$, and so are the asymptotic channel
wave functions. Neither the two-center Dirac equation, nor its boundary
conditions, are rigorously separable for finite $\gamma$.

In this context, we review the pioneering work of Eichler and co-workers
\cite{DeE94,Ei87,TE90} referred to by the name {\it Coulomb-boundary
conditions}, where the long-range Coulomb or Li\'enard-Wiechert
interaction was replaced by an effective short-range interaction.
We show how corrections of order $\gamma^{-2}$, explicit in our formal
definition of the short-range representation, are implicit in the
replacement procedure of the electron-projectile distance by the
  target-projectile distance that was used to obtain the asymptotic channels
with these Coulomb-boundary conditions.

In Sec.\ II, we discuss the asymptotic channel solutions
for the two-center Dirac equation for extremely relativistic
($\beta \rightarrow 1,~\gamma\rightarrow \infty$) heavy-ion collisions.
We derive factored forms for the asymptotic solutions which
are accurate to order $\gamma^{-2}$, i.e.\ they are exact in the
high-energy limit ($\gamma \rightarrow \infty$).
In Sec.\ II A,
we consider the case where the electron is asymptotically
referred to the target reference frame (i.e.\ the electron is near
to the target as $|t| \rightarrow \infty$), while
in Sec.\ II B,
we consider the case were the electron is asymptotically
near to neither the target nor the projectile ion, and is most
naturally referred to the collider (center-of-velocity) frame.
In Sec.\ III, we define and present the short-range
representation and derive from it
the high-energy or sharp limit for the two-center Dirac equation
in a simple form.
  In Sec.\ IV, we show that the phase transformation defining
the short-range representation constitutes a gauge transformation.
In so doing, we make explicit the connection between the
Coulomb-boundary conditions and the gauge transformations
first used by Baltz, Rhoades-Brown, and Weneser in numerically
solving the two-center Dirac equation via coupled-channel
methods\cite{BRW,BRW93,Ba95}.
Alternative treatments of the asymptotic electron-projectile distance
and alternative phase choices for the asymptotic channels are discussed
in the appendices.

\section{Asymptotic Solutions to Two-center Dirac Equation}
\label{as2cde}

We study relativistic heavy-ion collisions with
a single active electron, e.g.\ we neglect electron-electron
interactions in comparison to the strong electron-ion interactions.
An external-field approach to the influence of the ions
on the electron is appropriate for peripheral impact parameters,
heavy-ions, and high energies, where, to a very good approximation,
the ions travel on parallel, straight-line trajectories, and
ion recoil is negligible.
We are using natural units ($c=1$, $m_e=1$, and $\hbar=1$).
The quantity $\alpha$ is the fine-structure constant,
$\check{\alpha}$ and $\check{\gamma}^{\mu}$ are Dirac matrices
in the Dirac representation, as in Ref.~\cite{greiner},
and ${\rm I}_4$ is the 4-dimensional unit matrix.

\subsection{States referred to a target-fixed inertial frame}

Consider first a collision of a heavy, point-like projectile
ion having charge $Z_P$ with a target ion having charge $ Z_T$.
We consider the dynamics of a single electron interacting
with the external, time-dependent electromagnetic field
created by the two heavy ions (see Fig.\ \ref{targetframe}).
The position of the target nucleus is the origin of the
electron coordinates, and the electron has position vector
$ \vec{r}_T= \vec{r} = (x,y,z) $, and  time coordinate $t$.
The projectile moves with constant velocity, $\beta$,
parallel to the $z$ axis along a trajectory displaced from
the target by the impact parameter $\vec{b}$.
The projectile is located at the origin of the moving inertial frame,
and in the projectile frame the electron's position
vector is $ \vec{r}\,''_P = \vec{r}\,''= (x'',y'',z'') $, and
time coordinate $t''$.
Coordinates in the target and projectile inertial frames are related by an
inhomogeneous Lorentz transformation (Lorentz boost) parallel
to the $z$ axis such that
\begin{eqnarray}
  \vec{r}\,''_\perp &=& \vec{r}_\perp -\vec{b} \nonumber \\
  z'' &=& \gamma (z - \beta t)\; , \nonumber \\
  t'' &=& \gamma (t - \beta z)\; ,
\label{forwardLT}
\end{eqnarray}
where $ \vec{r}_\perp = (x,y)$ are the transverse spatial coordinates
of the electron in the target frame.
The Lorentz boost implies that the electron-projectile
distance in the projectile frame,
$ r''_P \equiv \sqrt{(x'')^2 + (y'')^2 + (z'')^2 }$, is represented in
target-frame coordinates as
\begin{equation}
  r''_P(\vec{r},t) = \sqrt{ (\vec{r}_\perp -\vec{b})^2 +
                     \gamma^2 (z - \beta t)^2 } \;.
\label{ePdistance}
\end{equation}
Equivalently, we may refer all coordinates to the projectile nucleus. The
resulting relations are obtained by the replacements $P\leftrightarrow T$,
$\beta\rightarrow -\beta$ and $\vec{b}\rightarrow -\vec{b}$.

\subsubsection{Two-center Dirac equation}

The single-center Dirac equation describing the bound
and continuum states of the target ion has the following
form in the target frame,
\begin{equation}
i\frac{\partial}{\partial t} |\psi_T(\vec{r},t)\rangle
 = \left[ \hat{H}_0  + \hat{H}_T    \right]
   |\psi_T(\vec{r},t)\rangle \; ,
\label{target}
\end{equation}
where
$ \hat{H}_0 $ is the free Dirac Hamiltonian, and
$ \hat{H}_T $ is the interaction of the electron with the target nucleus,
\begin{eqnarray}
\hat{H}_0 &\equiv& -i \check{\alpha}\cdot\vec{\nabla} + \check{\gamma^0} \;, \\
\hat{H}_T &\equiv&  - \frac{Z_T \alpha }{ r_T} \; .
\end{eqnarray}
By $\{ |\psi_T^{(j)}(\vec{r},t)\rangle \} $, we denote the
stationary states of the target ion
with quantum numbers $j$ (e.g.\ see for details Ref.\ \cite{EM95}).

The two-center, time-dependent Dirac equation in the target frame
for an electron interacting with both target and projectile
ions is
   \begin{equation}
i\frac{\partial}{\partial t} |\Psi(\vec{r},t)\rangle
 = \left[ \hat{H}_0  + \hat{H}_T + \hat{H}_P(t)
   \right] |\Psi(\vec{r},t)\rangle \;,
 \label{2centertarget}
\end{equation}
where $|\Psi(\vec{r},t)\rangle$ is the Dirac spinor wave
function of the electron, and
\begin{equation}
\hat{H}_P(t) \equiv
\frac{-Z_P \alpha \gamma ({\rm I}_4 - \beta\check{\alpha}_z)}
     {\sqrt{ (\vec{r}_\perp -\vec{b})^2 +  \gamma^2 (z - \beta t)^2 }}
\end{equation}
is the electron-projectile interaction.

\subsubsection{Coulomb-boundary conditions}

The interactions appearing in the two-center, time-dependent
Dirac equation, Eq.\ (\ref{2centertarget}),
are of long-range form, so that the distortion of the electron's
wavefunction induced by a distant ion should not, in principle, be
neglected \cite{Ei90,EM95,exception}.
Asymptotic channel wavefunctions are therefore defined
as the solution of the two-center Dirac equation for
asymptotic times.
The importance of including the electron's interaction with
asymptotically distant ions has been discussed extensively
by Eichler and coworkers \cite{DeE94,Ei87,TE90} for relativistic atomic
collisions in their work on the asymptotic
solutions known as the {\it Coulomb-boundary conditions}
(see Ref.\  \cite{EM95}, Sec.\ 5.3.3).

In defining the asymptotic channel solutions for the two-center
Dirac equation, Eq.\ (\ref{2centertarget}), the asymptotic
electron-projectile separation
$r''_P(\vec{r},t\rightarrow \infty)$ is approximated in Refs.\ \cite{EM95,TE90}
 by the internuclear separation $R''$
(see Appendix A, Eq.\ (\ref{A7})), that is
\begin{equation}\label{int1}
r''_P(\vec{r},|t|\rightarrow \infty) \rightarrow R'' =
  \sqrt{ b^2 + \gamma^2 (\beta^2z - \beta t)^2 } \;.
\end{equation}
This approximation transforms Eq.\ (\ref{2centertarget}) to the form
   \begin{equation}\label{int2}
i\frac{\partial}{\partial t} |\Phi^{R\infty}_T(\vec{r},t)\rangle
 = \left[ \hat{H}_0  + \hat{H}_T + \hat{H}^{R\infty}_P(t)
   \right] |\Phi^{R\infty}_T(\vec{r},t)\rangle \;,
\end{equation}
where $|\Phi^{R\infty}_T(\vec{r},t)\rangle$ is the asymptotic
solution, and
\begin{equation}\label{int3}
\hat{H}_P^{R\infty}(t) \equiv -
\frac{Z_P \alpha \gamma ({\rm I}_4 - \beta\check{\alpha}_z)}
     { \sqrt{ b^2 + \gamma^2 (\beta^2z - \beta t)^2 }} \; .
\end{equation}
is an {\it approximate} asymptotic electron-projectile interaction.

Equation (\ref{int2}) can be solved exactly for any value of $\beta$.
Consider an ansatz which is a product of a space-time dependent
phase factor and a single-center state (i.e.\ a function of the
electron-target distance),
\begin{equation}\label{int4}
 |\Phi_T^{R\infty}(\vec{r},t)\rangle
= e^{-i \chi^R_P(z,t)}
|\psi^{R\infty}(\vec{r},t)\rangle \; ,
\end{equation}
where the argument of the space-time dependent phase factor is
   \begin{eqnarray}\label{int5}
&&\chi^R_P(z,t) \equiv 
      \frac{Z_P\alpha}{\beta}\ln(R''-\beta t'')\nonumber\\
  &&=\frac{Z_P\alpha}{\beta} 
             \ln\left[\gamma (\beta^2 z-\beta t)
             + \sqrt{b^2 +\gamma^2(\beta^2 z-\beta t)^2}\right].
\end{eqnarray}
Substituting this ansatz into Eq.\ (\ref{int2}),
multiplying from the left by $ e^{i \chi^R_P(z,t)} $,
and collecting like terms gives
   \begin{equation}\label{int6}
i\frac{\partial}{\partial t}
|\psi^{R\infty}(\vec{r},t)\rangle
 = \left[ \hat{H}_0  + \hat{H}_T  \right]
|\psi^{R\infty}(\vec{r},t)\rangle  \;.
\end{equation}
With the ansatz (\ref{int4}), both the scalar and the vector components
of the asymptotic interaction (\ref{int3}) are canceled exactly,
and Eq.\ (\ref{int6}) is identical to Eq.\ (\ref{target}).
This means that
$|\Phi_T^{R\infty}(\vec{r},t)\rangle$ of Eq.\ (\ref{int4}) factors exactly
into a space-time dependent phase factor and a single-center target eigenstate
$|\psi^{R\infty} (\vec{r},t)\rangle = |\psi^{}_T(\vec{r},t)\rangle $.

The relativistic asymptotic
solutions of the form (\ref{int4}) are
exact only in the $\gamma \rightarrow
\infty$ limit.
For large, finite $\gamma$, the factored forms are very
useful, approximate asymptotic solutions  .

In the derivation reviewed here, the approximation occurs in
using Eq.\ (\ref{int1}) to obtain Eq.\ (\ref{int2}),
and not in the solution to  Eq.\ (\ref{int2}).
The  asymptotic distance, Eq.\ (\ref{int1}), is
accurate in the nonrelativistic limit $\beta^2\ll 1,\;\gamma\approx 1$
\cite{DeE94}, but becomes approximate for larger values of $\gamma$, when its
accuracy is of the order $\gamma^{-2}$ (see Appendix A).

\subsubsection{Asymptotic two-center Dirac equation}

Here we present an alternative derivation of the
factored asymptotic channel states. Formally, at the asymptotic limit,
Eq.~(\ref{2centertarget}) gives an asymptotic two-center Dirac
equation, (Eq.~(\ref{asy2centertarget}) below), that is {\it exact} in
the following sense: it is the rigorous mathematical limit of
Eq.~(\ref{2centertarget}) as $|t|\rightarrow\infty$.
We obtain this exact equation and then solve it {\it approximately},
to order $\gamma^{-2}$.

Consider again the case with the electron near to the
target at asymptotic times.
In this limit, the electron-projectile distance is (\ref{A11}),
\begin{equation}
\lim_{|t| \rightarrow \infty} r''_P(\vec{r},t)
  \equiv r''^{\infty}_P(\vec{r},t)
 = \sqrt{ b^2 + \gamma^2 (z - \beta t)^2 } \;.
\label{ePasymptotic}
\end{equation}
This expression differs from (\ref{ePdistance}) by neglecting the transverse
electron coordinate $\vec{r}^{}_{\perp}$, while the longitudinal coordinate $z$
is retained, since it enters into the Lorentz transformation (see Appendix A).
Using this distance to obtain the asymptotic limit of the
electron-projectile interaction,
the asymptotic, two-center Dirac equation in the target frame is
   \begin{equation}
i\frac{\partial}{\partial t} |\Phi^{\infty}_T(\vec{r},t)\rangle
 = \left[ \hat{H}_0  + \hat{H}_T + \hat{H}^{\infty}_P(t)
   \right] |\Phi^{\infty}_T(\vec{r},t)\rangle \;,
 \label{asy2centertarget}
\end{equation}
where $|\Phi^{\infty}_T(\vec{r},t)\rangle$ is the asymptotic
channel solution for an electron referred to the target frame,
and $  \hat{H}^\infty_P(t) $
is the exact asymptotic interaction of the electron with the distant
projectile,
\begin{equation}
\hat{H}_P^{\infty}(t) \equiv
\frac{-Z_P \alpha \gamma ({\rm I}_4 - \beta\check{\alpha}_z)}
     {\sqrt{ b^2 +  \gamma^2 (z - \beta t)^2 }} \; .
\end{equation}

For solutions to the asymptotic Dirac equation, Eq.\ (\ref{asy2centertarget}),
consider an ansatz which is a product of a space-time dependent
phase factor and a single-center state,
\begin{equation}
 |\Phi_T^{\infty}(\vec{r},t)\rangle
= e^{-i \chi^{}_P(z,t)}
|\psi^{\infty}(\vec{r},t)\rangle \; ,
\label{targetansatz}
\end{equation}
where the argument of the space-time dependent phase factor is
   \begin{eqnarray}
  \chi^{}_P(z,t) \equiv 
     \frac{Z_P\alpha}{\beta}
\ln\left[\gamma (z-\beta t)+\sqrt{b^2 +\gamma^2(z-\beta t)^2}\right] .
\label{target_phase}
\end{eqnarray}
Substituting this ansatz into Eq.\ (\ref{asy2centertarget}),
multiplying from the left by $ e^{i \chi^{}_P(z,t)} $,
and collecting like terms gives
   \begin{eqnarray}
&&i\frac{\partial}{\partial t}
|\psi^{\infty}(\vec{r},t)\rangle
  =  \left[ \hat{H}_0  + \hat{H}_T \right. \nonumber \\
 &&- \left. \left( \frac{1}{\gamma^2 -1} \right)
\frac{Z_P \alpha \gamma \beta\check{\alpha}_z}
     {\sqrt{ b^2 +  \gamma^2 (z - \beta t)^2 }}
   \right]
|\psi^{\infty}(\vec{r},t)\rangle  \;.
\label{dirac10}
\end{eqnarray}
The scalar component of the asymptotic electron-projectile
interaction is canceled exactly.
The remaining vector component is of order $1/\gamma^2$,
and vanishes in the $\gamma \rightarrow \infty$ limit.
In this limit,
the remaining equation is identical to the single-center
Dirac equation for the target ion, Eq.\ (\ref{target}),
and $|\psi^{\infty}(\vec{r},t)\rangle$ is therefore
a solution to this single-center
equation, $|\psi^{\infty}(\vec{r},t)\rangle
\rightarrow |\psi_T(\vec{r},t)\rangle$.
We conclude again that, in the extreme high-energy limit,
the solution to the asymptotic, two-center Dirac equation,
Eq.\ (\ref{asy2centertarget}), factors exactly into
an unperturbed, single-center target eigenstate,
$|\psi_T(\vec{r},t)\rangle$,
and a space-time dependent phase factor,
\begin{equation}
\lim_{\beta \rightarrow 1}  |\Psi_T^{\infty}(\vec{r},t)\rangle
= e^{-i \chi^{}_P(z,t)} |\psi_T(\vec{r},t)\rangle \; .
\label{Tasysolution}
\end{equation}

We have discussed two alternative derivations of the factored forms
for the asymptotic solutions for the two-center Dirac equation
and have shown that they provide identical results in the high-energy
limit: Equations (\ref{int2}) and (\ref{asy2centertarget}),
as well as their respective solutions,
Eqs.\ (\ref{int4}) and (\ref{Tasysolution}),
are identical as $\beta \rightarrow 1$.
The physical reason for this is simple. As $\beta\rightarrow 1$, the
target atom, as seen from the projectile,
shrinks to a disk, so that the distinction between the $z$-coordinate of the
nucleus and that of the electron disappears.

 For large, finite $\gamma$, both
derivations provide slightly different,
but equally useful, approximate solutions accurate to order $\gamma^{-2}$.
Other equally valid choices of the argument of the phase factor in
Eq.\ (\ref{target_phase}) can be made which differ only in factors
of $\beta^2$ \cite{Ei90,BRW} (see Appendix B).

\subsection{Collider frame}

For electrons distant from both the target and projectile ion
at asymptotic times, the collider (i.e.\ center-of-velocity)
inertial frame is a natural choice.
The origin of the collider frame is reached from the
origin of the target frame, for example, by an
inhomogeneous Lorentz transformation in the $z$ direction
to a frame of
velocity $ \beta_C = \sqrt{ 1-\gamma_C^{-2} } $
and Lorentz factor $\gamma_C = \sqrt{(\gamma+1)/2} $.
In the transverse direction, the origin of the collider frame
is located equidistant from the target and projectile trajectories
(see Fig.\ \ref{colliderframe}).
The position vector of the electron in the collider frame is
$ \vec{r}\,'_C = \vec{r}\,' = (x',y',z') $, and the
associated time is $ t'$.
Coordinates in the projectile and target frames are each related to
the coordinates in the collider frame by equal, but oppositely directed,
Lorentz transformations in the $z$ direction,
\begin{eqnarray}
  \vec{r}\,''_\perp &=& \vec{r}_\perp\,' - \vec{b}/2 \\
  z'' &=& \gamma_C (z' - \beta_C t')\; , \\
  t'' &=& \gamma_C (t' - \beta_C z')\; ,
\end{eqnarray}
and,
\begin{eqnarray}
  \vec{r}_\perp &=& \vec{r}_\perp\,' + \vec{b}/2 \\
  z &=& \gamma_C (z' + \beta_C t')\; , \\
  t &=& \gamma_C (t' + \beta_C z')\; .
\end{eqnarray}

As a consequence of the Lorentz boosts,
the electron-projectile distance
in collider-frame coordinates is
\begin{equation}
  r''_P(\vec{r}\,',t') = \sqrt{ (\vec{r}_\perp\,' -\vec{b}/2)^2 +
                     \gamma^2_C (z' - \beta_C t')^2 } \;,
\end{equation}
and the electron-target distance
in collider-frame coordinates is
\begin{equation}
  r_T(\vec{r}\,',t') = \sqrt{ (\vec{r}_\perp\,' + \vec{b}/2 )^2 +
                     \gamma^2_C (z' + \beta_C t')^2 } \;.
\end{equation}

\subsubsection{Two-center Dirac equation}

The free-particle Dirac equation in the collider frame
 has the form
\begin{equation}
i\frac{\partial}{\partial t'} |\phi'_C(\vec{r}\,',t')\rangle
 =  \hat{H}_0'  |\phi'_C(\vec{r}\,',t')\rangle \; ,
\label{free}
\end{equation}
where
$\hat{H}_0'$ is the free Dirac Hamiltonian in the collider frame,
\begin{equation}
\hat{H}'_0 \equiv -i \check{\alpha}\cdot\vec{\nabla}' + \check{\gamma^0} \; .
\end{equation}
The set $\{ |\phi'^{(j')}_C(\vec{r}\,',t')\rangle \} $ represents the Dirac
plane-wave eigenstates with quantum numbers $j'$, namely,
the three components of the momentum, $\vec{j}$,
the sign of the energy, and the spin.

The two-center, time-dependent Dirac equation in the collider frame
for an electron interacting with both target and projectile
ions is
   \begin{equation}
i\frac{\partial}{\partial t'} |\Psi'(\vec{r}\,',t')\rangle
 = \left[ \hat{H}'_0
        + \hat{H}_T'(t')
        + \hat{H}_P'(t')
   \right] |\Psi'(\vec{r}\,',t')\rangle \;,
 \label{2centercollider}
\end{equation}
where $|\Psi'(\vec{r}\,',t')\rangle$ is the Dirac spinor wave
function of the electron,
$ \hat{H}_T'(t') $ is the electron-target interaction,
and $ \hat{H}_P'(t') $ is the electron-projectile interaction,
\begin{equation}
\hat{H}_T'(t') \equiv
\frac{-Z_T \alpha \gamma_C ({\rm I}_4 + \beta_C\check{\alpha}_z)  }
     {\sqrt{ (\vec{r}_\perp\,' + \vec{b}/2)^2
              + \gamma_C^2 (z' + \beta_C t')^2 }} \; ,
\end{equation}
\begin{equation}
\hat{H}_P'(t') \equiv
\frac{-Z_P \alpha \gamma_C ({\rm I}_4 - \beta_C\check{\alpha}_z)}
     {\sqrt{ (\vec{r}\,'_\perp -\vec{b}/2)^2
              + \gamma_C^2 (z' - \beta_C t')^2 }} \; .
\end{equation}

\subsubsection{Asymptotic two-center Dirac equation}

Consider, in the collider frame, at asymptotic times,
an electron distant from both the target
and projectile ions. The electron-projectile and electron-target
distances then have the following asymptotic limits,
\begin{eqnarray}
\lim_{|t'| \rightarrow \infty} r''_P(\vec{r}\,',t')
&\equiv& r''^\infty_P(\vec{r}\,',t') \nonumber \\ 
&=&
  \sqrt{(b/2)^2 + \gamma^2_C (z' - \beta_C t')^2 }  \;,
\nonumber \\
\lim_{|t'| \rightarrow \infty} r_T(\vec{r}\,',t')
&\equiv& r_T^\infty (\vec{r}\,',t') \nonumber \\
&=&
  \sqrt{(b/2)^2 + \gamma^2_C (z' + \beta_C t')^2 } \; .
\label{PTasymptotic}
\end{eqnarray}
Using these distances, the asymptotic, two-center Dirac
equation is
   \begin{eqnarray}
&& i\frac{\partial}{\partial t'} |\Phi'^{\infty}_C(\vec{r}\,',t')\rangle
 =  \nonumber \\
&& \left[ \hat{H}'_0
        + \hat{H}_T'^\infty(t')
        + \hat{H}_P'^\infty(t')
   \right] |\Phi'^\infty_C(\vec{r}\,',t')\rangle \;,
 \label{asy2centercollider}
\end{eqnarray}
where $|\Phi'^\infty_C(\vec{r}\,',t')\rangle$ is the Dirac spinor wave
function of the electron
asymptotic channel solution,
$ \hat{H}_T'^\infty(t') $ is the asymptotic electron-target interaction,
and $ \hat{H}_P'^\infty(t') $ is the asymptotic electron-projectile
interaction,
\begin{equation}
\hat{H}_T'^\infty(t') \equiv
\frac{-Z_T \alpha \gamma_C ({\rm I}_4 + \beta_C\check{\alpha}_z)  }
     {\sqrt{ (b/2)^2  + \gamma_C^2 (z' + \beta_C t')^2 }} \; ,
\end{equation}
\begin{equation}
\hat{H}_P'^\infty(t') \equiv
\frac{-Z_P \alpha \gamma_C ({\rm I}_4 - \beta_C\check{\alpha}_z)}
     {\sqrt{ (b/2)^2 + \gamma_C^2 (z' - \beta_C t')^2 }} \; .
\end{equation}

For the solutions of Eq.\ (\ref{asy2centercollider}),
consider an ansatz of a space-time dependent phase factor
times a Dirac plane-wave state.
\begin{equation}
 |\Phi_C'^{\infty}(\vec{r}\,',t')\rangle
= e^{-i \chi_C'(z',t')}
|\phi'^{\infty}(\vec{r}\,',t')\rangle \; ,
\label{collideransatz}
\end{equation}
where
   \begin{eqnarray}
&& \chi_C'(z',t') \equiv \nonumber \\
&&
\frac{Z_P\alpha}{\beta}
\ln\left[\gamma_C (z-\beta_Ct')
+\sqrt{(b/2)^2 +\gamma_C^2(z'-\beta_Ct')^2}\right]
\nonumber \\
&-&
\frac{Z_T\alpha}{\beta}
\ln\left[ \gamma_C (z+\beta_Ct')
+\sqrt{(b/2)^2 +\gamma_C^2(z'+\beta_Ct')^2} \right] \;.
\label{colliderphase}
\end{eqnarray}
Substituting Eq.\ (\ref{colliderphase}) into Eq.\ (\ref{asy2centercollider}),
multiplying from the left by $ e^{+i \chi_C'(z',t')} $, and
collecting like terms gives
   \begin{eqnarray}
&&i\frac{\partial}{\partial t'}
|\phi'^{\infty}(\vec{r}\,',t')\rangle = \nonumber \\
 && \left[ \hat{H}'_0  +
\left( \frac{1}{\gamma^2 -1} \right)
\frac{Z_T \alpha \gamma \beta_C\check{\alpha}_z}
     {\sqrt{ (b/2)^2 + \gamma_C^2 (z' + \beta_C t')^2 }} \right.
 \\
&-&
\left.
\left( \frac{1}{\gamma^2 -1} \right)
\frac{Z_P \alpha \gamma \beta_C\check{\alpha}_z}
     {\sqrt{ (b/2)^2 + \gamma_C^2 (z' - \beta_C t')^2 }}
   \right]
|\phi'^{\infty}(\vec{r}\,',t')\rangle . \nonumber
\end{eqnarray}
As in the target-centered case,
the scalar component of the asymptotic electron-projectile
and electron-target interactions cancel exactly, and
the vector component vanishes in the $\beta_C \rightarrow 1 $ limit.
In this limit,
the remaining equation is identical to the free Dirac equation,
Eq.\ (\ref{free}), and
$|\phi'^{\infty}(\vec{r}\,',t')\rangle
\rightarrow |\phi_C'(\vec{r}\,',t')\rangle$, is a Dirac plane-wave
eigenstate.
We conclude that in the extreme, high-energy limit,
the ansatz in Eq.\ (\ref{collideransatz})
with the Dirac plane wave,
is the exact solution
to the asymptotic, two-center Dirac equation,
Eq.\ (\ref{asy2centercollider}),
\begin{equation}
\lim_{\beta_C \rightarrow 1} |\Phi_C'^{\infty}(\vec{r}\,',t')\rangle
= e^{-i \chi_C'(z',t')} |\phi_C'(\vec{r}\,',t')\rangle \; .
\label{Casysolution}
\end{equation}

\section{Short-range Representation}
\label{srr}

The factored forms of the asymptotic solutions to the two-center
Dirac equation, Eqs.\ (\ref{int4},\ref{Tasysolution},\ref{Casysolution}),
obtained in the previous section,
invite the definition of a new
representation for the time-dependent Dirac equation.
In this section, we introduce this representation, which we
call the {\em short-range representation}, within
the context of computing amplitudes for direct reactions first in the
target frame, and then the collider frame.

In nonrelativistic \cite{DeE94} as well as in relativistic collisions
\cite{Ei87,TE90}, it has been previously
shown to be useful   to introduce a
formulation that  substitutes the long-range Coulomb or Li\'enard-Wiechert
interaction by an effective short-range interaction,
jointly with an appropriate phase transformation,
thus rendering formal scattering theory applicable.
The essence of these approaches has been to replace the
electron-projectile separation for an electron close to the target
and asymptotically far from the projectile, by the internuclear
separation $R''$ given by the expression (\ref{int1}).
Then, with an ansatz like Eq.\ (\ref{int4}), the approximate
asymptotic electron-projectile interaction (\ref{int3}) can be removed
completely from the Hamiltonian, so that for {\em finite}
electron-projectile separations, one has to deal with a
{\em short-range interaction} obtained from the original
long-range one by the replacement
\begin{equation}
\label{int7}
    \frac{1}{r''_P}\rightarrow \frac{1}{r''_P} - \frac{1}{R''}.
\end{equation}
The effects of subtracting the asymptotic long-range part have been
demonstrated numerically for direct and rearrangement collisions
using perturbation theory and coupled-channel methods \cite{TE90}.

We have shown   in the previous section and in appendix A
that in the relativistic regime $R''$ differs from a more
rigorous asymptotic limit
for the electron-projectile separation (\ref{ePasymptotic}) or (\ref{A11}),
 by terms of the order of $1/\gamma^2$.
This approach revealed that a complete and exact removal of the asymptotic
electron-projectile interaction is possible only in the
$\beta \rightarrow 1$ limit (see Eqs.\ (\ref{dirac10}) and (\ref{B3})).

For finite relativistic energies,
terms of the order $1/\gamma^2$ remain in
either the scalar or vector components of the electron-projectile
  asymptotic interaction, but are small for large $\gamma$.
In the following, we are concentrating on
  the high-energy limit, in which the description becomes simple and unique.

\subsection{Exact Transition Amplitudes}

Following the notation of
Ref.\ \cite{Ei90},
let $ |\Psi^{(+)}_j (t_f)\rangle $ be the exact outgoing-wave
solution evolving from an
initial channel solution $ |\Phi_j^{\infty} (t_i) \rangle $, i.e.\
\begin{equation}
\lim_{t \rightarrow - \infty}  |\Psi^{(+)}_j (t)\rangle
=  |\Phi_j^{\infty} (t) \rangle \; ,
\end{equation}
and $ |\Phi_k^{\infty} (t_f) \rangle $ be the final asymptotic
channel.
Then, by definition, the {\em exact} transition amplitude
is given in the {\it post} form as
\begin{equation}
A^{(+)}_{kj}=\lim_{t_f\rightarrow\infty}
\langle\Phi^{\infty}_k (t_f)|\Psi^{(+)}_j (t_f)\rangle \; .
\label{post}
\end{equation}
The {\em prior} form of the amplitude is defined at $ t \rightarrow -\infty$
as the projection of the exact incoming wave solution
$ |\Psi^{(-)}_j (t_i)\rangle $ evolving backward in time from
the final channel  $ |\Phi_k^{\infty} (t_f) \rangle $, i.e.\
\begin{equation}
\lim_{t \rightarrow  \infty}  |\Psi^{(-)}_k (t)\rangle
=  |\Phi_k^{\infty} (t) \rangle \; ,
\end{equation}
onto the initial channel solution  $ |\Phi_j^{\infty} (t_i) \rangle $,
\begin{equation}
A^{(-)}_{kj}=\lim_{t_i \rightarrow -\infty}
\langle\Psi^{(-)}_k (t_i)|\Phi^{\infty}_j (t_i)\rangle \; .
\label{prior}
\end{equation}

The post and prior forms of the amplitude may be
unified using the time-evolution operator $ \hat{U}(t_f,t_i) $
to relate the full outgoing-wave (incoming-wave) solution to its
initial (final) state as
\begin{eqnarray}
 |\Psi^{(+)}_j (t_f)\rangle &=& \hat{U}(t_f,t_i)
 |\Phi_j^{\infty} (t_i) \rangle \nonumber  \\
 |\Psi^{(-)}_k (t_i)\rangle &=& \hat{U}^\dagger(t_f,t_i)
 |\Phi_k^{\infty} (t_f) \rangle \; .
\label{tevolve}
\end{eqnarray}
Inserting Eqs.\ (\ref{tevolve}) into Eq.\ (\ref{post}) or
Eq.\ (\ref{prior}), one obtains,
\begin{equation}
A_{kj}=
\lim_{ \stackrel{t_i\rightarrow -\infty}{t_f\rightarrow \infty} }
\langle\Phi^{\infty}_k (t_f)|\hat{U}(t_f,t_i)
   | \Phi^{\infty}_j (t_i) \rangle \;.
\label{amplitude}
\end{equation}
  Reference \cite{Ei90} considered these states
in the target inertial frame.
Yet, the definitions presented here
apply to the projectile or collider frame as well.
In direct reactions, the initial and final channels
in Eq.\ (\ref{amplitude})
are both solutions of the same asymptotic Hamiltonian
  associated with a single collision partner
(e.g.\ atomic excitation or ionization).
In rearrangement collisions, the initial and final channels may be
solutions of different asymptotic Hamiltonians
associated with different collision partners (e.g. charge exchange).

\subsection{Short-range Dirac equation}

In this section, we discuss the short-range representation for the
Dirac equation within the context of computing transition amplitudes
for {\it direct reactions} in the high-energy limit.

\subsubsection{Equation of motion: target frame}

In the following, we consider the limit
$\beta \rightarrow 1 $, so that the asymptotic channels for
a target-frame electron interacting with a nearby target ion
and a distant projectile ion has the exact, factored solution of
Eq.\ (\ref{Tasysolution}).
We substitute this asymptotic solution into the expression for
the exact transition amplitudes for direct reactions in the target frame,
Eq.\ (\ref{amplitude}), for the initial state $j$ and final state $k$,
\begin{eqnarray}
&&A_{kj}= \\
&&\lim_{ \stackrel{t_i\rightarrow -\infty} {t_f\rightarrow \infty} }
\langle e^{-i \chi^{}_P(z,t_f)} \psi^{(k)}_T (t_f)|
\hat{U}(t_f,t_i)
   |e^{-i \chi^{}_P(z,t_i)} \psi^{(j)}_T (t_i) \rangle \;.
\nonumber
\end{eqnarray}
Rearranging the exponential factors in the expression so that
they are applied directly to the evolution operator, one obtains
\begin{eqnarray} \label{amplitudeb1}
&&A_{kj}=   \\
&&\lim_{ \stackrel{t_i\rightarrow -\infty} {t_f\rightarrow \infty} }
\langle  \psi^{(k)}_T (t_f)|
e^{+i \chi^{}_P(z,t_f)} \hat{U}(t_f,t_i)
   e^{-i \chi^{}_P(z,t_i)} | \psi^{(j)}_T (t_i) \rangle \;.
\nonumber
\end{eqnarray}

The transition amplitude, Eq.\ (\ref{amplitudeb1}), is suggestive
of a new representation
for the dynamics through the operation of the space-time-dependent
phase,
\begin{eqnarray}
&& | \Psi^{(S)} (\vec{r},t) \rangle \equiv
e^{+i \chi^{}_P(z,t)}  | \Psi (\vec{r},t) \rangle \; ,
\label{srwf}
\\
&& \hat{U}^{(S)}(t_f,t_i)   \equiv e^{+i \chi^{}_P(z,t_f)}  \hat{U}(t_f,t_i)
e^{-i \chi^{}_P(z,t_i)} \; ,
\label{srteo}
\end{eqnarray}
where $ | \Psi^{(S)} (\vec{r},t) \rangle $ is the wavefunction,
and $\hat{U}^{(S)}(t_f,t_i)$ is the time-evolution operator
in the new representation.
Substituting Eq.\ (\ref{srteo}) into Eq.\ (\ref{amplitudeb1}) gives
the exact amplitude for direct reactions in the new representation,
\begin{equation}
A_{kj}=
\lim_{ \stackrel{t_i\rightarrow -\infty} {t_f\rightarrow \infty} }
\langle  \psi^{(k)}_T (t_f)|
 \hat{U}^{(S)}(t_f,t_i)
   | \psi^{(j)}_T (t_i) \rangle \;.
\label{sramplitude}
\end{equation}
Note that Eq.\ (\ref{sramplitude}) has the form of a transition
amplitude computed between initial and final channels which are undistorted
single-center eigenstates of the target ion, as would be the case
if the interaction between the electron and the distant projectile
was of short range.

To understand better its utility,
we transform the two-center Dirac equation into the short-range representation.
Beginning with Eq.\ (\ref{2centertarget}), and making the substitution
\begin{equation}
| \Psi (\vec{r},t) \rangle =
e^{-i \chi^{}_P(z,t)}
| \Psi^{(S)} (\vec{r},t) \rangle\;,
\end{equation}
gives, after multiplying from the left by $ e^{+i\chi^{}_P(z,t)} $,
the equation of motion,
   \begin{equation}
 i\frac{\partial}{\partial t} |\Psi^{(S)}(\vec{r},t)\rangle
 = \left[ \hat{H}_0  + \hat{H}_T + \hat{W}_P(t)
   \right] |\Psi^{(S)}(\vec{r},t)\rangle \;,
\label{sr2centertarget}
\end{equation}
where $ \hat{W}_P(t) $ is the time-dependent electron-projectile
interaction in the new representation \cite{Ba95},
\begin{equation}
\hat{W}_P(t) \equiv  \hat{H}_P(t) -
\frac{-Z_P \alpha \gamma ({\rm I}_4 - (1/\beta)\check{\alpha}_z)}
     {\sqrt{ b^2 +  \gamma^2 (z - \beta t)^2 }} \; .
\end{equation}
In the high-energy limit, $ \beta \rightarrow 1 $, and
\begin{equation}
\lim_{\beta \rightarrow 1} \hat{W}_P(t)
\equiv \hat{H}_P(t) - \hat{H}^\infty_P(t) \; .
\end{equation}
$ \hat{W}_P(t) $ is the original electron-projectile
interaction with its long-range, asymptotic space-time dependence
subtracted.
The cancellation is exact only in the $\beta \rightarrow 1 $ limit.
Otherwise, there remains a residual long-range
interaction of the order $1/\gamma^2$.
As a result of this very useful characteristic,
we name this new representation the {\em short-range representation}.
The phase transformation used to define the short-range
representation, Eq.\ (\ref{srwf}), exactly cancels the
phase distortion factor contained in the asymptotic solution
to the two-center Dirac equation in the extreme, high-energy
limit, Eq.\ (\ref{Tasysolution}).
The result is a representation of the two-center Dirac equation
appropriate for direct reactions in extremely relativistic
heavy-ion collisions in which the electron-projectile interaction
has short range and the initial and final states are effectively
single-center eigenstates of the target ion.
(Note that the transverse-coordinate dependence of $ \hat{W}_P(t) $ remains
of long-range (i.e.\ $1/r_\perp$) form.
However, the transverse coordinates do not contribute to the interaction
of the electron with a distant ion at asymptotic times.)

The electron-ion interaction in the short-range representation
simplifies further if, in addition to the $\beta \rightarrow 1$ limit,
one requires that the transverse electron coordinates
$\vec{r}_\perp$ and the impact parameter $b$ are small
compared to $\gamma$, i.e.\
\begin{equation}
| \vec{r}_\perp |,\, b \ll \gamma\;.
\label{sharpelimit}
\end{equation}
In this limit, $ \hat{W}_P(t) $ factors into
a product of a Dirac-delta function of argument $(t-z)$
and a logarithmic function of the transverse coordinates
(similar to the potential induced by a line of charge),
(see Refs. \cite{Ja92,Ba97,SW98}), i.e.
\begin{equation}
\lim_{ \stackrel{ \beta \rightarrow 1 }{ r_\perp, b \ll \gamma } }
\hat{W}_P(t) =
({\rm I}_4 - \check{\alpha}_z ) Z_P \alpha \delta(t-z)
\ln\left[ \frac{(\vec{r}_\perp - \vec{b})^2  }{b^2  }   \right] \;.
\label{Wsharpelimit}
\end{equation}
We refer to this as the {\em sharp limit} of the electron-projectile
interaction in the short-range representation,
as the interaction has zero range in the light-front coordinate
$ \tau_- \equiv (t-z)/2$.
This behavior reflects the fact that the peak transverse electric field
produced by a moving charge increases proportional to $\gamma$ while the
duration $\Delta t \approx b/(\gamma \beta)$ of the collision decreases as
$ 1/ \gamma$.
The interaction in this sharp limit has the character of an
electromagnetic shockfront which develops as the speed of the
source of the electromagnetic field, $\beta$, approaches the propagation
speed, $c$, of the field \cite{footnote1}.

The short-range, two-center Dirac equation, Eq.\ (\ref{sr2centertarget}),
in the sharp limit, (i.e.\ using the interaction in
Eq.\ (\ref{Wsharpelimit})),
has been recently used
by Baltz
to compute the high-energy limit
of the   impact-parameter dependent probabilities for  bound-free
electron-positron pair
production in peripheral, relativistic heavy-ion collisions \cite{Ba97}.
In reflecting on this achievement, it is important to recall that
the derivation of Eq.\ (\ref{sr2centertarget}) given here
assumes asymptotic channels which correspond to {\it direct reactions
only}. Asymptotic channels which correspond to the electron being
distant from the target as either $t_i \rightarrow -\infty$ or
$ t_f \rightarrow +\infty$ are not considered in this description.
As a result, the {\it charge-transfer mechanism} for bound-free
pair production \cite{Ei95,IE96} is not included in the solutions
given in Ref.\ \cite{Ba97}.
The extreme high-energy behavior of the charge-transfer mechanism
for pair production has not received detailed study.

An analogous short-range representation may be defined for direct reactions
in the projectile frame, with similar interpretation.
The construction of the short-range representation in the collider
frame is also similar, but differs in that the asymptotic interaction
of the electron with both projectile and target ions must be
considered. We discuss the collider-frame case in the next section.

\subsubsection{Equation of motion: collider frame}

Consider the extreme, high-energy limit
$\beta_C \rightarrow 1 $ of the two-center Dirac equation in
the collider frame, Eq.\ (\ref{2centercollider}), so that the
asymptotic channels for
an electron interacting with  distant target and projectile ions
has the factored form of Eq.\ (\ref{Casysolution}).
We substitute this exact solution into the expression for
the exact transition amplitudes for the collider frame
for the initial state $j$ and final state $k$,
\begin{eqnarray}
A_{kj}=
\lim_{ \stackrel{t'_i\rightarrow -\infty} {t'_f\rightarrow \infty} }
&&
\langle e^{-i \chi'_C(z',t'_f)} \phi'^{(k)}_C (t'_f)|  \\
&&
\times \hat{U}'(t'_f,t'_i)
   |e^{-i \chi'_C(z',t'_i)} \phi'^{(j)}_C (t'_i) \rangle \;.
\nonumber
\end{eqnarray}
Rearranging the exponential factors in the expression so that
they are applied directly to the evolution operator, one obtains,
\begin{eqnarray}
\label{Camplitudeb1}
A_{kj}=
\lim_{ \stackrel{t'_i\rightarrow -\infty} {t'_f\rightarrow \infty} }
&&
\langle  \phi'^{(k)}_C (t'_f)|  \\
&&\times e^{+i \chi'_C(z',t'_f)} \hat{U}'(t'_f,t'_i)
   e^{-i \chi'_C(z',t'_i)} | \phi'^{(j)}_C (t'_i) \rangle \;.
\nonumber
\end{eqnarray}
Defining the short-range representation in the collider frame,
\begin{eqnarray}
&& | \Psi'^{(S)} (\vec{r}\,',t') \rangle \equiv
e^{+i \chi'_C(z',t')}  | \Psi' (\vec{r}\,',t') \rangle
\label{Csrwf}
\\
&& \hat{U}'^{(S)}(t'_f,t'_i) \equiv e^{+i \chi'_C(z',t'_f)}
\hat{U}'(t'_f,t'_i)
e^{-i \chi'_C(z',t'_i)} \; .
\label{Csrteo}
\end{eqnarray}
gives the formal expression for the exact transition amplitude
between plane-wave states in the collider frame using the short-range
representation,
\begin{equation}
A_{kj}=
\lim_{ \stackrel{t'_i\rightarrow -\infty} {t'_f\rightarrow \infty} }
\langle  \phi'^{(k)}_C (t'_f)|
 \hat{U}'^{(S)}(t'_f,t'_i)
   | \phi'^{(j)}_C (t'_i) \rangle \;.
\label{Csramplitude}
\end{equation}

To obtain the two-center Dirac equation in the collider frame
in the short-range representation, we
begin with Eq.\ (\ref{2centercollider}), and make the substitution
\begin{equation}
| \Psi'(\vec{r}\,',t') \rangle =
e^{-i \chi'_C(z',t')}
| \Psi'^{(S)} (\vec{r}\,',t') \rangle\;.
\end{equation}
After multiplying from the left by $ e^{+i\chi'_C(z',t')} $,
the equation of motion
has the form
   \begin{eqnarray}
\label{sr2centercollider}
&& i\frac{\partial}{\partial t} |\Psi'^{(S)}(\vec{r}\,',t')\rangle
 = \nonumber \\
&& \left[ \hat{H}'_0  + \hat{W}'_T(t') + \hat{W}'_P(t')
   \right] |\Psi'^{(S)}(\vec{r}\,',t')\rangle \; ,
\end{eqnarray}
where $ \hat{W}'_T(t') $ and  $ \hat{W}'_P(t') $
are the
time-dependent electron-target and electron-projectile interactions
in the  short-range representation,
\begin{eqnarray}
\hat{W}'_T(t') \equiv  \hat{H}'_T(t') -
\frac{-Z_T \alpha \gamma_C ({\rm I}_4 +
(1/\beta_C)\check{\alpha}_z)}
     {\sqrt{ (b/2)^2 +  \gamma^2 (z' + \beta_C t')^2 }} \;,
\nonumber \\
\hat{W}'_P(t') \equiv  \hat{H}'_P(t') -
\frac{-Z_P \alpha \gamma_C ({\rm I}_4 -
(1/\beta_C)\check{\alpha}_z)}
     {\sqrt{ (b/2)^2 +  \gamma^2 (z' - \beta_C t')^2 }} \; .
\end{eqnarray}
In the high-energy limit, $\beta_C \rightarrow 1$, and
\begin{eqnarray}
\lim_{\beta_C \rightarrow 1} \hat{W}'_T(t')
&\equiv& \hat{H}'_T(t') - \hat{H}'^\infty_T(t') \; ,
\nonumber \\
\lim_{\beta_C \rightarrow 1} \hat{W}'_P(t')
&\equiv& \hat{H}'_P(t') - \hat{H}'^\infty_P(t') \; .
\label{srWcollider}
\end{eqnarray}
As in Eq.\ (\ref{sr2centertarget}), the asymptotic  dependence of the
time-dependent interaction has been canceled exactly (in the $\beta
\rightarrow 1$ limit).
Likewise,
the phase distortion in the asymptotic
channel solutions is canceled by the phase transformation
defining the  short-range representation,
and the asymptotic channels are effectively the Dirac plane
waves.

Applying the sharp limit of Eq.\ (\ref{sharpelimit}) to
Eqs.\ (\ref{srWcollider}), we obtain  the following
factored forms for the time-dependent interaction
\cite{Ja92,Ba97,SW98},
\begin{eqnarray}
&&\lim_{ \stackrel{ \beta_C \rightarrow 1 }{ r'_\perp, b \ll \gamma_C } }
 \hat{W}'_T(t')= \nonumber \\
 &&
({\rm I}_4 + \check{\alpha}_z ) Z_T \alpha \delta(t'+z')
\ln\left[ \frac{(\vec{r}\,'_\perp + \vec{b}/2)^2  }{(b/2)^2} \right] \;,
\nonumber \\
&&\lim_{ \stackrel{ \beta_C \rightarrow 1 }{ r'_\perp, b \ll \gamma_C } }
\hat{W}'_P(t')=  \nonumber \\
&&
({\rm I}_4 - \check{\alpha}_z ) Z_P \alpha \delta(t'-z')
\ln\left[ \frac{(\vec{r}\,'_\perp - \vec{b}/2)^2  }{(b/2)^2} \right] \;.
\label{sharpcollider}
\end{eqnarray}
The short-range, two-center Dirac equation, Eq.\ (\ref{sr2centercollider}),
in the sharp limit (using Eqs.\ (\ref{sharpcollider})) has recently
been used to compute the high-energy limit of the free electron-positron
pair-production amplitudes in peripheral relativistic
heavy-ion collisions \cite{SW98,BM98,SW98b}.
As in the case of the target frame equation considered previously,
the amplitudes derived from Eq.\ (\ref{sr2centercollider}) and given in
Refs.\ \cite{SW98,BM98,SW98b,ER98a} correspond to direct reactions only.
For the present case in the collider frame, only asymptotic
electron states distant from both target and projectile ions
are considered \cite{footnote2}.
The contribution of other asymptotic channels to the high-energy
limit of free-pair production requires further investigation.

\section{Gauge transformations}

In discussing the two-center Dirac equation in the target frame,
Eq.\ (\ref{2centertarget}), for relativistic heavy-ion collisions,
Baltz and coworkers have regarded
the phase transformation, Eq.\ (\ref{srwf}),
used here to define the short-range representation, as a gauge
transformation \cite{BRW,BRW93,Ba95,Ba97}.
Eichler {\it et al.}\ have also remarked
that the phase factors obtained
in solving for the asymptotic channel solutions of Eq.\
(\ref{2centertarget})
and used to obtain a short-range effective interaction can be interpreted
as gauge transformations \cite{EM95,Ei90}.
In this section, we show explicitly that the phase transformation used
to define the short-range representation is equivalent to a gauge
transformation, and highlight the relatedness of these two
viewpoints.

In investigating the phase transformation, Eq.\ (\ref{srwf}), as a
gauge transform, it is convenient to write the two-center Dirac
equation explicitly in terms of
the electromagnetic four-vector interaction $A^\mu$.
Beginning with Eq.\ (\ref{2centertarget}), we write the electron-projectile
interaction Hamiltonian as $ \hat{H}_P(t) = A_0 - \check{\alpha}_z  A_z $,
where
\begin{eqnarray}
A_0(\vec{r},t) &\equiv& \frac{  - Z_P \alpha \gamma }{ r''_P(\vec{r},t) } \;,
\nonumber \\
A_z(\vec{r},t) &\equiv& \beta A_0(\vec{r},t) \;,
\end{eqnarray}
so that the two-center Dirac equation is written in the form
   \begin{equation}
i\frac{\partial}{\partial t} |\Psi(\vec{r},t)\rangle
 = \left[ \hat{H}_0  + \hat{H}_T + A_0(t) - \check{\alpha}_z  A_z(t)
   \right] |\Psi(\vec{r},t)\rangle \;.
 \label{2centertarget_A}
\end{equation}
We now re-derive the two-center Dirac equation in the short-range
representation, Eq.\ (\ref{sr2centertarget}), by substituting
the phase transformation in Eq.\ (\ref{srwf}) into
Eq.\ (\ref{2centertarget_A}) and
multiplying from the left by $ e^{-i \chi_P}$, to obtain
   \begin{eqnarray}
&& i\frac{\partial}{\partial t} |\Psi^{(S)}(\vec{r},t)\rangle
 = \nonumber \\
&& \left[ \hat{H}_0  + \hat{H}_T
+ {\rm I}_4 A^{(S)}_0(t) - \check{\alpha}_z  A^{(S)}_z(t)
   \right] |\Psi^{(S)}(\vec{r},t)\rangle \;,
 \label{srg2centertarget}
\end{eqnarray}
where the components of the four-vector interaction in the
short-range representation are
\begin{eqnarray}
A_0^{(S)}(\vec{r},t) \equiv A_0(\vec{r},t)
                      - \frac{ \partial \chi_P (z,t) }{ \partial t} \;,
\nonumber \\
A_z^{(S)}(\vec{r},t) \equiv A_z(\vec{r},t)
                      + \frac{ \partial \chi_P (z,t) }{ \partial z} \;,
\label{gtpots}
\end{eqnarray}
or, more explicitly,
\begin{eqnarray}
A_0^{(S)}(\vec{r},t)
 &=& -Z_P \alpha \gamma \left[ \frac{1}{ r''_P(\vec{r},t)  }
                            - \frac{1}{ r''^{\infty}_P(\vec{r},t)  } \right]
\nonumber \\
A_z^{(S)}(\vec{r},t)
 &=& -Z_P \alpha \gamma \beta \left[ \frac{1}{ r''_P(\vec{r},t)  }
                            - \frac{1/\beta^2}{ r''^{\infty}_P(\vec{r},t)  }
\right] \; .
\label{srpots}
\end{eqnarray}
With the interaction written in the form of
Eqs.\ (\ref{gtpots}), the phase transformation in
Eq.\ (\ref{srwf}) clearly accomplishes a gauge transformation.

In general, gauge transformations leave physical quantities,
such as the S-matrix amplitudes, invariant, whereas other
quantities, such as wavefunctions, propagators, and asymptotic channel
solutions, may depend on the gauge.
Clearly, the invariance of physical quantities relies on an exact
formulation. From a practical point of view, however, approximations are often
needed. A widely applied method consists in the expansion of the time-dependent
wave function in terms of a basis set of channel functions, such that the time
dependent Dirac equation (\ref{2centertarget}) is equivalent to an infinite set
of coupled equations for the time-dependent expansion coefficients. For
practical reasons, this set is truncated at a finite number of states. While
the complete set is, of course, invariant under gauge transformations, a finite
set usually is not. In fact, a gauge transformation may not only modify the
effective interaction, but it also affects the convergence property of the
expansion \cite{EiB96}. Therefore, both effects should always be considered
simultaneously and, actually, can be utilized to speed up convergence
\cite{TE90}.

Within an exact treatment, which is our main subject, a distinction has
been made in Refs.\
\cite{BRW,Ru90,Ru93} between gauge transformations which leave the asymptotic
channels invariant (or trivially modified) as a result
of the gauge function being constant at asymptotic times,
and those which modify boundary conditions since the
gauge function is not constant asymptotically.
Indeed, the gauge transformation considered
in defining the short-range representation modifies the
asymptotic states since it behaves
asymptotically as
\begin{eqnarray}
\lim_{t \rightarrow + \infty} \chi_P (z,t)
&=& \frac{-Z_P \alpha}{\beta} \ln \left[
\frac{ 2 \gamma |z-\beta t| }{ b^2}
\right]  \; ,
\\
\lim_{t \rightarrow - \infty}\chi_P (z,t)
&=& \frac{+Z_P \alpha}{\beta} \ln \left[
 2 \gamma |z-\beta t|
\right]  \;.
\end{eqnarray}

Implicit in using the short-range representation (or gauge)
in the high-energy limit is that the phase
transformation, Eq.\ (\ref{srwf}),
defining the representation, exactly cancels the
phase distortion of the asymptotic channels induced by the distant
projectile ion (see Eq.\ (\ref{Tasysolution})).
As a result, the asymptotic channel solutions for direct reactions
in the short-range representation are the undistorted, single-center
atomic states, $ | \psi_j (\vec{r},t) \rangle $.
In other words, by using undistorted atomic states as asymptotic
channels in the short-range representation, as was done
in Ref.\ \cite{Ba97}, one is, in effect, using the
factored form for the asymptotic channel solutions,
Eq.\ (\ref{Tasysolution}), of
Eichler and coworkers.

\section{Conclusions}

A primary goal of this work was to place on a
clear and firm theoretical foundation the ``Sharp Dirac equation'',
i.e. the two-center Dirac equation(s), in both the target and the
collider frames, in the short-range representation, in the extreme
relativistic (sharp) limit.
The reason this is of primary importance is that the extreme
relativistic limit of the two-center Dirac equation in the 
short-range representation
for heavy-ion collisions simplifies remarkably,
and allows for closed form solutions for pair-production
amplitudes in this limit.

With these goals in mind, we have described the relationship between
asymptotic solutions to the two-center, time-dependent Dirac equation
for a single electron in peripheral relativistic heavy-ion collisions,
and phase (or gauge) transformations designed to remove
the long-range asymptotic interaction from the equation of motion.
 Direct reactions are central to the discussion.
``Charge-transfer'' mechanisms for pair production \cite{Ei95,IE96}
have been omitted here, and should be subsequently considered
in the high-energy limit.

We have shown that the asymptotic channel solutions factorize into a
space-time dependent phase and an eigenstate of the appropriate
time-independent Hamiltonian,
in the limit $ \beta \rightarrow 1 $.
For collision velocities less than the speed of light, this
factorization is approximate with accuracy of the order $1/\gamma^2$.
We have also shown that as a result of this
factorization a gauge transformation may be performed
to a new representation in which the asymptotic
dynamics are included in the states.
In this representation,
the asymptotic interaction between the electron and
a distant ion is of short-range form, and the asymptotic
solutions are undistorted, stationary solutions of
a time-independent Hamiltonian.
Under such conditions, a formally correct formulation
of scattering theory may be constructed.
In addition, this short-range representation has advantages
for the convergence of numerical
calculations\cite{Ei87,TE90,BRW,BRW93,BRW94}.

The factorization of the asymptotic solutions in the
$\beta \rightarrow 1$ limit also
provides a significant simplification
in the dynamics.
A further simplification is achieved if the magnitude
of the transverse coordinate, $r_\perp$, and the
impact parameter, $b$, are constrained to be much smaller
than $\gamma$.
In this limit, the time-dependent interaction factors into
a logarithmic function of the transverse coordinate, and
a Dirac-delta function of a light-front variable,
$\tau_\pm = (z \pm t)/2$,  describing an electromagnetic
shock on the lightfront \cite{Ja92,Ba95,SW98}.
The identification of the separable form has allowed for the
closed-form solution of amplitudes for electron-positron pair
production in the high-energy limit of heavy-ion
collisions\cite{Ba97,SW98,BM98,SW98b}.

We have also made a connection with the previous pioneering work of
Eichler {\it et al.}\ on the Coulomb-boundary conditions.
We have elucidated and discussed the relatedness of the Coulomb-boundary
approach and what Baltz and coworkers have recently accomplished
via the machinery of gauge-transformations.
We have shown that in the high energy limit, these two approaches
are 
in agreement, and differ mostly in their language.

The replacement strategy previously developed by Eichler and coworkers
was designed to remove the long-range part by a gauge or phase
transformation. 
This treatment is fully symmetric with respect to the
target and projectile frame, has the correct nonrelativistic limit
and has been successful in a number of calculations.
We note however, that there
is no unique way to derive ``replacements''.
When the purpose is to have a good basis-set for
numerical calculation, a replacement procedure
is a useful and adequate approximation.
On the other hand, if one is treating the problem in a formal 
approach, as was recently done
for the high-energy collision limit
in Refs.\  \cite{Ba97,SW98,BM98,SW98b},
the rigorous definition of the short-range representation as presented
here is of   significant importance.

In regard to using the factored solution as an
accurate, but approximate, asymptotic channel solution for calculation
of high-energy collision phenomena,
one should keep in mind that the factored solution, e.g.\ the phase factor
times a single-center eigenstate, is an exact solution to the
two-center Dirac equation for asymptotically large times, {\it only} in
the limit $\beta\rightarrow 1$.
For large, but finite $\gamma$, the factored solution is an approximate
solution to the asymptotic Dirac equation accurate to order $1/\gamma^2$.
Hence, choosing between
factors of $\beta$ in the argument of the phase is largely a matter of
personal taste.

\section*{Acknowledgments}

This work was supported by
the Center for Computational Sciences Division of Oak Ridge
National Laboratory, Lockheed-Martin Energy Research under
contract DE-AC05-96OR22464 with the U.S. Department of Energy,
and by the National Science Foundation through a grant for the
Institute for Theoretical Atomic and Molecular Physics at Harvard
University and Smithsonian Astrophysical Observatory.

\appendix

\section{Asymptotic electron-ion distance}
\label{asydistance}

In this appendix, we discuss the asymptotic limit of
the electron-projectile distance needed to describe the
interaction of an electron with a projectile ion
at asymptotic times $|t|\rightarrow \infty$.
The electron-projectile distance in the projectile's rest frame,
\begin{equation}
 r''_P = \sqrt{ (\vec{r}\,''_\perp)^2 + (z'')^2 } \;,
\label{A1}
\end{equation}
is represented in terms of the target-frame coordinates as
\begin{equation}
 r''_P(\vec{r},t)
 = \sqrt{ (\vec{r}_\perp - \vec{b} )^2 + \gamma^2 (z-\beta t)^2 }\;,
\label{A2}
\end{equation}
where $\{\vec{r}\,''_\perp, z'', t''\}$ and $\{\vec{r}_\perp , z, t\}$
are the space-time coordinates of the electron
in the projectile frame and in the target frame, respectively,
which are related by an inhomogeneous Lorentz transformation
as in Eqs.\ (\ref{forwardLT}).
We would like to obtain an asymptotic (i.e. $|t| \rightarrow \infty$)
limit for $r_P''$ when the internuclear
separation $R''$,
\begin{equation}\label{A4}
R''(t'') = \sqrt{ b^2 + \beta^2 (t'')^2 } \; ,
\end{equation}
is large compared to the separations between
the electron and the target.

We now discuss the problem of the asymptotic electron-projectile separation in
two different versions.

\subsubsection{Internuclear separation}

Following the arguments given in Refs.\ \cite{Ei90,EM95},
we {\it substitute} the internuclear separation $R''$ for the asymptotic
electron-projectile separation, which should be a good approximation for very
large positive or negative times. Formally, this corresponds to taking
\begin{equation}
\vec{r}_\perp \rightarrow 0\;, \;\;
z \rightarrow 0
\label{A5}
\end{equation}
in the target frame. Once this {\em replacement} is performed, we consider
$R''$ as a {\em parameter of the system} describing the internuclear motion and
no longer  the position of the electron, that is, we leave the Lorentz
transformation,
\begin{equation}
t'' = \gamma ( t - \beta z )\;,
\label{A6}
\end{equation}
relating the projectile-frame time to the target-frame time for an arbitrary
electron position, intact. This  constitutes an inconsistency, if the
position of the target nucleus with respect to the projectile nucleus is
interpreted
as an electronic position with the coordinates (\ref{A5}). In this respect, the
replacement $\vec{r}\,''_P\rightarrow R''$ with $R''$ given by Eqs.\ (\ref{A4})
and
(\ref{A6}), i.e.
\begin{equation}
r''_P \rightarrow
\sqrt{ b^2 + \gamma^2 \left( \beta^2 z - \beta t \right)^2 }
\label{A7}
\end{equation}
is not the {\em formally derivable} asymptotic limit. According to Eq.\
(\ref{A5}),
{\em formal consistency} can only be achieved by replacing $t''\rightarrow
\gamma t$, i.e.
\begin{equation}
r''_P \rightarrow \sqrt{ b^2 + \gamma^2 \beta^2 t^2 } \; .
\label{A8}
\end{equation}
However, Eq.\ (\ref{A8}) is not useful for our purpose, because the
$z$-dependence is required to describe the magnetic component of the
electromagnetic interaction at asymptotic times for large $\gamma$ \cite{Ei87}.

In the nonrelativistic limit $\beta^2\ll 1, \;\gamma\approx 1$,
when $t''\approx t$, no inconsistency occurs, and Eqs.\
(\ref{A4}) or (\ref{A7}) immediately lead to the usual and
successfully applied replacement \cite{DeE94}
\begin{equation}
\label{A9}
\vec{r}^{}_P\rightarrow R=\sqrt{b^2+\beta^2 t^2}.
\end{equation}

\subsubsection{Longitudinal electron-projectile separation}

To obtain a formally rigorous $z$ dependence for the asymptotic limit
of the electron-projectile distance $r''_P$ for an electron near
to the target but distant from the projectile,
one should not use $\vec{r}\,''_P\rightarrow R''$, but should maintain
the dependence on the $z$ coordinate, that is, retain the exact longitudinal
electron-projectile distance. Even if $|z|\ll \beta |t|$ in the laboratory at
asymptotic times, we do not set it to zero. This means that instead of Eq.\
(\ref{A5}), we take
 \begin{equation}
\label{A10}
  \vec{r}_{\perp}\rightarrow 0\;,
\end{equation}
while $z$ is retained.
This procedure guarantees that Lorentz transformations can be consistently
applied. The resulting, {\em formally correct} asymptotic limit of the
electron-projectile distance is
\begin{equation}
\lim_{|t|\rightarrow \infty} r''_P (\vec{r},t)
\equiv r''^\infty_P(\vec{r},t) =
\sqrt{ b^2 + \gamma^2 \left(z - \beta t \right)^2 } \; .
\label{A11}
\end{equation}
This, no doubt, is a better approximation to Eq.\ (\ref{A2}) than
 $\vec{r}\,''_P\rightarrow R''$.
In order to compare it with Eq.\ (\ref{A7}), we
may write
\begin{equation}
\lim_{|t|\rightarrow \infty} r''_P (\vec{r},t) =
\sqrt{ b^2 + \gamma^2 \left[ (\gamma^{-2} + \beta^2) z - \beta t
\right]^2 } \; .
\end{equation}
Note that, compared to Eq.\ (\ref{A7}), an additional term with $1/\gamma^2$
appears. This term reflects the difference between taking the longitudinal
electron-projectile separation and the internuclear separation.
  One sees this difference more explicitly by considering the ratio
$ r''_P/R''$ in the limit $\gamma \gg 1$, $|z| \ll \beta |t|$,
and $ b \ll \gamma |t|$. Keeping terms proportional to $z/t$, we
obtain
\begin{equation}
\frac{r''_P }{ R''} \approx 1 - \frac{ z}{\gamma^2 t} \;.
\end{equation}
Indeed, for
very large values of $\gamma$, the target atom as seen from the projectile
shrinks to a disk, so that the electronic $z$-coordinate almost coincides with
the $z=0$ coordinate of an electron located at the target nucleus.

We here have discussed two different
approaches for identifying the
asymptotic
electron-projectile separation. The first is based on a substitution by the
internuclear separation, which implies a formal
 inconsistency if interpreted as a true {\em electronic} separation instead of
a parameter describing the projectile motion. However, it appears
physically
reasonable and has the correct nonrelativistic limit. The second is formally
rigorously derivable by keeping the longitudinal electronic coordinate and
hence encounters no problems when applying Lorentz transformations in a
straightforward fashion.
Both approaches differ in a term of the order of $1/\gamma^2$ in the asymptotic
electron-projectile separation and agree for $\beta\rightarrow 1$.
As discussed in Sec.\ II A and in Appendix \ref{choices},
discrepancies of this order propagate into the factored forms of
the asymptotic channel solutions and the asymptotic interaction
when they are applied for large, but finite, $\gamma$.

\section{Phase choices for asymptotic channels}
\label{choices}

In Sec.\ II A, we have discussed two versions , Eqs.\ (\ref{int4}) and
(\ref{targetansatz}), for separating asymptotic wave functions by introducing
the phases (\ref{int5}) and (\ref{target_phase}), respectively. These phases,
differing only in factors $\beta^2$, arise from different choices
for the asymptotic electron-ion distance (see Appendix \ref{asydistance}).
For our present purposes,
choosing among these different phase arguments is largely
a matter of personal taste since only in the $\gamma \rightarrow \infty$
limit does the asymptotic interaction vanish exactly in the
short-range representation. For large finite values of $\gamma$, terms of the
order of $1/\gamma^2$ remain. In order to illustrate the consequences of phase
choices, consider yet another product ansatz
 for the solution of the asymptotic
two-center Dirac equation in the target frame,
\begin{equation}
 |\Phi_T^{\infty}(\vec{r},t)\rangle
= e^{-i \Lambda_P(z,t)} |\psi_T(\vec{r},t)\rangle \; ,
\label{targetansatz2}
\end{equation}
where
   \begin{equation}
 \Lambda_P(z,t) \equiv
Z_P\alpha \beta
\ln\left[\gamma (z-\beta t)+\sqrt{b^2 +\gamma^2(z-\beta t)^2}\right]
 \;.
\label{target_phase2}
\end{equation}
Substituting this ansatz into Eq.\ (\ref{asy2centertarget}),
multiplying from the left by $ e^{i \Lambda_T(z,t)} $,
and collecting like terms gives
   \begin{eqnarray}\label{B3}
&&i\frac{\partial}{\partial t} |\psi_T(\vec{r},t)\rangle
 = \left[ \hat{H}_0  + \hat{H}_T \right. \nonumber \\
&& -
\left.
\left( \frac{1}{\gamma^2} \right)
\frac{Z_P \alpha \gamma {\rm I}_4}
     {\sqrt{ b^2 +  \gamma^2 (z - \beta t)^2 }}
   \right] |\psi_T(\vec{r},t)\rangle \;.
\end{eqnarray}
With the phase choice in Eq.\ (\ref{target_phase2}),
the vector component of the asymptotic electron-projectile
interaction is canceled exactly,
and the remaining scalar component is of the order $1/\gamma^2$.
In contrast, with the phase choice made in Eq.\ (\ref{target_phase}),
which differs from Eq.\ (\ref{target_phase2}) only by a factor of $\beta^2$,
the scalar component cancels exactly, and the vector component
is of order $1/\gamma^2$.

One may always perform a gauge transformation such that a single
component (or a single linear combination of components) of the
four-vector electromagnetic interaction is exactly zero for all
times.
Such a gauge condition is known as an {\em axial gauge} (see
Refs.\ \cite{greiner,WO92}).
The novelty of the short-range representation
in the $\beta \rightarrow 1$ limit is that in it the full, asymptotic
interaction (both scalar and vector components) is zero.



\pagebreak

\begin{figure}[h]
\caption{
Coordinate systems for a relativistic collision between two ions.
The position of the target ion, with charge $ Z_T$, is the origin
of the unprimed coordinates.
The position of the projectile ion, with charge $ Z_P$, is the origin
of the   doubly primed coordinates. The projectile moves with constant
velocity $ \beta $ parallel to the $z$ axis on a trajectory with
impact parameter $ \vec{b}$.
The electron $ e^-$ has the coordinate $ \vec{r}_T $ with respect to
the target frame and $ \vec{r}\,''_P$ with respect to the projectile
frame.
}
\label{targetframe}
\end{figure}

\begin{figure}[h]
\caption{
Coordinate systems for a relativistic collision between to ions
similar to Fig.\ \protect\ref{targetframe}\protect\ except that
the collider (or center-of-velocity) frame is shown in addition.
The electron has the coordinates $ \vec{r}\,'_C $ with respect to
the collider frame.
The projectile and target ions have the collider-frame
coordinates, $ \vec{R}'_P $, and $ \vec{R}'_T $, respectively.
}
\label{colliderframe}
\end{figure}


\begin{thebibliography}{99}
\bibitem{exception}
       Exceptions to this rule include the case of screening
       of the long-range Coulomb interactions, which is not
       included in our idealized three-body system.
\bibitem{footnote1}
The high-energy limit of the electron-projectile interaction may also
be represented, in an alternative gauge, using a Heavyside step function
\cite{Ja92,SW98,BM98} which more directly displays its shockfront character.
\bibitem{footnote2}
  The restriction of considering only direct reactions is discussed
in Refs.\ \cite{SW98,SW98b} in terms of obtaining solutions only
``off the lightfronts'', and in Ref.\ \cite{BM98} by considering
only ``centrally produced'' electrons and positrons.
These two restrictions, as discussed in the references,
possess the same basic physical content.
\end{thebibliography}
\end{document}